\documentclass[aps,pra,floatfix,reprint,showpacs,amsmath,amssymb,
superscriptaddress]{revtex4-1}
\usepackage{graphicx}
\usepackage{hyperref}
\begin{document}

\title{Reflection and Refraction of the Laser Light Pulse
at the Vacuum--Medium Interface}

\author{W \.Zakowicz}
\email[]{zakow@ifpan.edu.pl}
\affiliation{Institute of Physics, Polish Academy of Sciences,
Al.~Lotnik\'ow 32/46, 02--668 Warsaw, Poland}

\author{A. A. Skorupski}
\email[]{andrzej.skorupski@ncbj.gov.pl}
\affiliation{Department of Theoretical Physics, National Centre for Nuclear
Research, Ho\.za 69, 00--681 Warsaw, Poland}

\date{\today}

\begin{abstract}
By generalizing the well known results for reflection and refraction of plane
waves at the vacuum--medium interface to Gaussian light beams, we obtain
analytic formulas for reflection and refraction of the TM and TE laser light
pulses. This enables us to give a possible explanation why no reflection was
observed in light pulse photographs in some vicinity of the air--resin
interface, given in L. Gao, J. Liang, C. Li, and L. V. Wang, Nature
\textbf{516}, 74 (2014). We suggest how to modify the experimental setup so as
to observe the reflected pulse.
\end{abstract}


\maketitle

\section{Introduction}\label{intro}

In an impressive paper \cite{gao} describing ultrafast photography ($10^{11}$
frames per second), among examples given there was a laser light pulse
photograph in some vicinity of the vacuum--resin interface, see Fig.~3(b) in
\cite{gao}. The transmitted pulse could be seen there, but no reflected one
was observed. Here we try to explain why there was no reflection,
and how to modify the experimental setup to observe it.

In Sec.~\ref{boundc} we show that the Gaussian beams for which the transverse
dimensions are constant along the beam, behave in full analogy to plane waves,
and we derive formulas pertaining to their reflection and refraction at
the plane vacuum--medium interface.

Section \ref{laserp} deals with reflection and refraction of laser pulses,
and Sec.~\ref{discus} specifies these results to the pulses used in
\cite{gao} and contains the final conclusions.

We assume that in our laboratory Cartesian coordinate system ($x,y,z$), with a
medium at $y<0$ and vacuum at $y>0$, we can fulfil the boundary conditions at
the interface, $y=0$, by a superposition of three beams propagating in the
$x,y$ plane: the incident (i), reflected (r) and refracted (or transmitted, t)
Gaussian beam.
In the paraxial approximation, we can describe any such steady state TM
beam as \cite{goldsm,scheck} (Gaussian units)
\begin{subequations}\label{gaussb1}
\begin{eqnarray} 
H_{z_{\text{i,r,t}}} &=& C_{\text{i,r,t}} \exp \biggl(
-\frac{x^2_{\text{i,r,t}}}{w_{x_{\text{i,r,t}}}^2}  -
\frac{z^2}{w_z^2} \biggr)\nonumber\\
&&\times\exp \bigl\{ i \bigl[ (k y)_{\text{i,r,t}} -
\omega t \bigr] \bigr\},\\
E_{x_{\text{i,r,t}}} &=& -Z_{\text{i,r,t}} H_{z_{\text{i,r,t}}},
\qquad
k_{\text{i,r,t}} = \omega \, n_{\text{i,r,t}}/c,
\end{eqnarray}
\end{subequations}
where $C_{\text{i}} = 1$, $C_{\text{r}} = R_{\text{TM}}$ is the reflection
coefficient, $C_{\text{t}} = T_{\text{TM}}$ is the transmission coefficient,
$y_{\text{i,r,t}}$ are Cartesian coordinates along each beam, $x_{\text{i,r,t}}$
are similar coordinates for transverse directions, $\omega$ is the angular
frequency, $c$ is the speed of light in vacuum, $k_{\text{i,r,t}}$
are the wave numbers, $n_{\text{i,r}} = 1$ and
$n_{\text{t}} = n \equiv \sqrt{\epsilon\mu}$ are the refraction indexes of
vacuum and medium, $Z_{\text{i,r}}=1$ and $Z_{\text{t}}=\sqrt{\mu/\epsilon}$
are similar field impedances, and the positive constants $\epsilon$ and $\mu$
characterize the medium.

Note that for each beam, the radii $w_x$ and $w_z$ are constant here
($y$ independent), which requires $y$ to be small as compared to the Rayleigh
ranges $\hat{y}_x$ and $\hat{y}_z$:
\begin{equation}
(y/\hat{y}_{x,z})^2 \ll 1, \qquad \hat{y}_{x,z} =  \pi w_{x,z}^2/\lambda,
\label{cond}
\end{equation}
where $\lambda = (c/n)2\pi/\omega$ is the wavelength.

Note that by TM (or TE) polarization we mean orthogonality of the
\textbf{H} (or \textbf{E}) vector with respect to the plane of incidence
($x$, $y$ in our case).

\section{Boundary conditions}\label{boundc}

In the limit $w_{x} \to \infty$ and $w_{z} \to \infty$ in (\ref{gaussb1}),
we obtain the plane waves: $H_z = C\exp\bigl[i(ky - \omega t)\bigr]$.
Therefore in this limit, the continuity of the tangential components of
\textbf{E} and \textbf{H} for the incident plus reflected beam versus the
transmitted one at the interface ($y = 0$) implies the Snell law
\begin{equation}
\varphi_{\text{r}} = \varphi_{\text{i}}, \qquad
n \sin\varphi_{\text{t}} = \sin\varphi_{\text{i}},
\end{equation}
and the Fresnel formulas \cite{scheck,jackson} $\bigl(\nu = n
\cos\varphi_{\text{t}}\bigr)$
\begin{subequations}
\begin{align}
R_{\text{TM}}& = \frac{\epsilon \cos\varphi_{\text{i}} - \nu}{\epsilon
\cos\varphi_{\text{i}} + \nu},& T_{\text{TM}}& =
1 + R_{\text{TM}}, \\
R_{\text{TE}}& = \frac{\mu \cos\varphi_{\text{i}} - \nu}{\mu
\cos\varphi_{\text{i}} + \nu},& T_{\text{TE}}& =
1 + R_{\text{TE}},\label{RTEM}
\end{align}
\end{subequations}
where $\varphi_{\text{i}}$, $\varphi_{\text{r}}$ and $\varphi_{\text{t}}$ are
the angles of incidence, reflection and refraction.

For finite radii $w_{x_{\text{i,r,t}}}$ and $w_{z}=w_{z_{\text{i,r,t}}}$, we
have to take into account linear relations defining $x_{\text{i,r,t}}$ and
$y_{\text{i,r,t}}$ as functions of the laboratory $x$ and $y$
($z_{\text{i,r,t}}=-z$):
\begin{equation}
x_{\text{i,r,t}} = (x,y)\cdot\hat{\mathbf{x}}_{\text{i,r,t}}, \qquad
y_{\text{i,r,t}} = (x,y)\cdot\hat{\mathbf{y}}_{\text{i,r,t}},\label{xyirt}
\end{equation}
where hats denote unit vectors for local coordinate axes of each beam,
\begin{equation}
\hat{\mathbf{x}}_{\text{i,r}} = (\pm\cos\varphi_{\text{i}},
\sin\varphi_{\text{i}}), \qquad
\hat{\mathbf{x}}_{\text{t}} = (\cos\varphi_{\text{t}},
\sin\varphi_{\text{t}}),
\end{equation}
\begin{equation}
\hat{\mathbf{y}}_{\text{i,r}} = (\sin\varphi_{\text{i}},
\mp\cos\varphi_{\text{i}}), \qquad
\hat{\mathbf{y}}_{\text{t}} = (\sin\varphi_{\text{t}},
-\cos\varphi_{\text{t}}).\label{haty}
\end{equation}
At the boundary, $y=0$, equations (\ref{xyirt}) lead to
\begin{equation}
\frac{x^2_{\text{i,r}}}{w_{x_{\text{i,r}}}^2} =
x^2\frac{\cos^2\varphi_{\text{i}}}{w_{x_{\text{i,r}}}^2}, \qquad
\frac{x^2_{\text{t}}}{w_{x_{\text{t}}}^2} =
x^2\frac{\cos^2\varphi_{\text{t}}}{w_{x_{\text{t}}}^2}.\label{cond1}
\end{equation}
Thus if $x=z=0$, Eq.~(\ref{gaussb1}) is the same as that for plane waves, and
the boundary conditions are satisfied. They will be satisfied also for nonzero
$x$ and $z$, if the RHSs of (\ref{cond1}) are the same, i.e., if
\begin{equation}
w_{x_{\text{r}}} = w_{x_{\text{i}}}, \qquad w_{x_{\text{t}}} =
w_{x_{\text{i}}}
\frac{\cos\varphi_{\text{t}}}{\cos\varphi_{\text{i}}} > w_{x_{\text{i}}}.
\label{wxt}
\end{equation}

For the TE Gaussian beam one has to replace $\text{TM} \to \text{TE}$,
$H_{z_{\text{i,r,t}}} \to E_{z_{\text{i,r,t}}}$, and $H_{x_{\text{i,r,t}}} =
E_{z_{\text{i,r,t}}}/Z_{\text{i,r,t}}$.

\section{Specification to laser pulses}\label{laserp}

If the incident laser beam is not monochromatic, but has a spectrum function
$F(\omega)$, we have to include all harmonics, and the time behavior of all
beams will be given by the integral (Inverse Fourier Transform)
\begin{equation}
\int_{-\infty}^{\infty} F(\omega)
\exp (i\omega \bar{t})\,d\,\omega \qquad \text{at} \qquad  \bar{t}=
\frac{y_{\text{i,r,t}}}{c/n_{\text{i,r,t}}} - t  .\label{int}
\end{equation}
For a Gaussian spectrum function
\begin{equation}
F(\omega) = \exp\{-[(\omega - \omega_0)/b]^2\},\label{F}
\end{equation}
centered about $\omega = \omega_0$, the Gaussian wave packets will be
obtained. In that case, for a conveniently normalized TM laser beam,
we end up with (neglecting the common factor $b\sqrt{\pi}$):
\begin{widetext}
\begin{eqnarray}
H_z(y>0) &=& \exp\biggl[ -\frac{x_{\text{i}}^2}{w_{x_{\text{i}}}^2} -
\frac{(y_{\text{i}} - c t)^2}{w_{y_{\text{i}}}^2} - \frac{z^2}{w_z^2} +
i k_0 (y_{\text{i}} - c t) \biggr] + R_{\text{TM}} \exp\biggl[
- \frac{x_{\text{r}}^2}{w_{x_{\text{i}}}^2} - \frac{(y_{\text{r}}- c t)^2}%
{w_{y_{\text{r}}}^2} - \frac{z^2}{w_z^2}
+ i k_0 (y_{\text{r}} - c t)\biggr],\ \ \label{Hza}\\
H_z(y \leq 0) &=& (1 + R_{\text{TM}})\exp\biggl[
- \frac{x_{\text{t}}^2}{w_{x_{\text{t}}}^2} - \frac{(y_{\text{t}}- c t/n)^2}%
{w_{y_{\text{t}}}^2} - \frac{z^2}{w_z^2}
+ i k_0 n (y_{\text{t}} - c t/n)\biggr],\label{Hzb}
\end{eqnarray}
\end{widetext}
where
\begin{equation}
w_{y_{\text{i}}} = 2c/b, \qquad w_{y_{\text{r}}} = w_{y_{\text{i}}}, \qquad
w_{y_{\text{t}}} = w_{y_{\text{i}}}/n,\label{wyrt}
\end{equation}
are the the light pulse radii along each beam,
while $k_0=2\pi/\lambda_0$, $\lambda_0=c T_0$ and $T_0=2\pi/\omega_0$ are
the vacuum wavenumber, wavelength and period for the beam carrier,
$\omega=\omega_0$. See also (\ref{xyirt})--(\ref{haty}) and (\ref{wxt}) for
remaining definitions, in which $\varphi_{\text{t}} =
\arcsin[(\sin\varphi_{\text{i}})/n]$.

Replacing $H_z \to E_z$ and $R_{\text{TM}} \to R_{\text{TE}}$ in
(\ref{Hza}) and (\ref{Hzb}), we obtain
the TE beam fields.

The real parts of the complex coordinates $H_z$ and $E_z$ defined above give
us the $z$ coordinates of the real physical fields.

\begin{figure}
\includegraphics[scale=0.5]{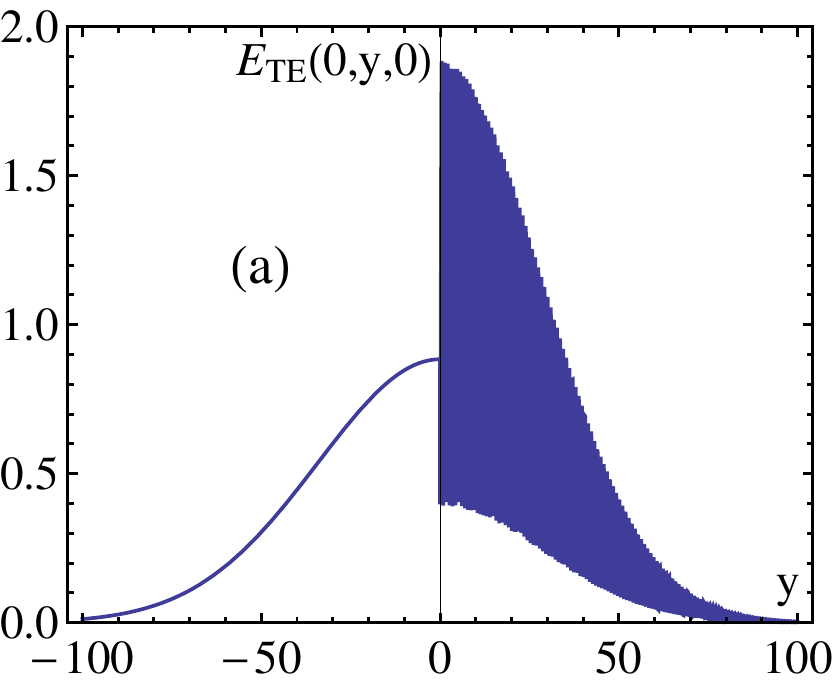}\includegraphics[scale=0.5]{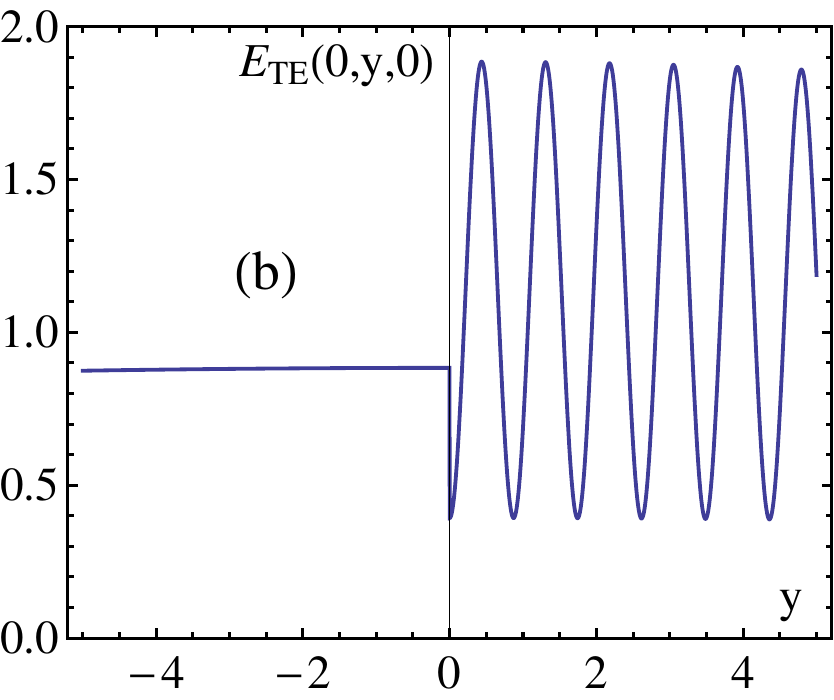}
\caption{(Color online) (a): $\mathcal{E}_{\text{TE}}(x=0,y,t=0)$ versus $y$.
(b): Blowup demonstrating discontinuity in $\mathcal{E}_{\text{TE}}$  due to
discontinuity in $\epsilon$ at the interface ($y=0$),
$\mathcal{E}_{\text{TE}}(y=0^-) = \epsilon\,\mathcal{E}_{\text{TE}}(y=0^+)$.
This also demonstrates continuity of $E_z$ at the interface.}
\label{ydep}
\end{figure}

From now on, it will be convenient to work with dimensionless variables,
by measuring time and space dimensions in the units of $T_0$ and
$\lambda_0$, see Figs.~\ref{ydep}--\ref{intpeaks}. This implies $c=1$ and
$k_0=2\pi$ in (\ref{Hza}) and (\ref{Hzb}).

At $t=0$, for each of the pulses in (\ref{Hza}) and (\ref{Hzb}),
the real part is a product
of a strongly oscillating function of $y_{\text{i,r,t}}$
[$\cos(2\pi y_{\text{i}})$, $\cos(2\pi y_{\text{r}})$, and $\cos(2\pi n
y_{\text{t}})$, with wavelengths $\lambda_{\text{i}} =
\lambda_{\text{r}} = 1$, $\lambda_{\text{t}} = 1/n$], and the envelope
which is a Gaussian proportional to
\begin{equation}
\exp\bigl[ -(\text{space coordinate}/\text{pulse radius})^2 \bigr].
\label{envel}
\end{equation}
During time evolution starting at some $t=t_{\text{in}}<0$,
the incident and reflected pulses move along the $y_{\text{i}}$
and $y_{\text{r}}$ axes with unit velocities, and the refracted pulse
along the $y_{\text{t}}$ axis with the velocity $1/n$.

Notice that there is a common factor
$\exp\bigl(-z^2/w_z^2\bigr)$ in (\ref{Hza}) and (\ref{Hzb}). Its square
will appear as a common factor in formulas defining energy densities
(time averaged over fast oscillations with $\omega=2\omega_0$), proportional
to $\mu H_z H_z^*$ and $\epsilon E_z E_z^*$.
Integrating these formulas $dz$ from $-\infty$ to
$\infty$ we obtain (time averaged) surface energy densities,
$\mathcal{E}_{\text{TM}}(x,y,t)$ and $\mathcal{E}_{\text{TE}}(x,y,t)$.
Again neglecting the common factor ($w_z\sqrt{\pi/2}$), we end up with
\begin{widetext}
\begin{eqnarray}
\mathcal{E}_{\text{TM}}(x,y>0,t) &=& \exp\biggl[-\Bigl(\frac{x_{\text{i}}}
{w_{x_{\text{i}}}/\sqrt{2}}\Bigr)^2 -
\Bigl(\frac{y_{\text{i}} - t}{w_{y_{\text{i}}}/\sqrt{2}}\Bigr)^2
\biggr] + R_{\text{TM}}^2 \exp\biggl[
- \Bigl(\frac{x_{\text{r}}}{w_{x_{\text{i}}}/\sqrt{2}}
\Bigr)^2 - \Bigl(\frac{y_{\text{r}}- t}{w_{y_{\text{i}}}/\sqrt{2}} \Bigr)^2
\biggr]\nonumber\\
&&+ 2 R_{\text{TM}} \exp\Biggl[- \frac{x_{\text{i}}^2 + x_{\text{r}}^2}
{w_{x_{\text{i}}}^2} -
\frac{(y_{\text{i}} - t)^2 + (y_{\text{r}} - t)^2}{w_{y_{\text{i}}}^2}
\Biggr] \cos\bigl(4\pi y \cos\varphi_{\text{i}}\bigr),\label{ETMir}\\
\mathcal{E}_{\text{TM}}(x,y \leq 0,t) &=& \mu (1 + R_{\text{TM}})^2
\exp\biggl[-\Bigl(\frac{x_{\text{t}}}{w_{x_{\text{t}}}/\sqrt{2}}\Bigr)^2 -
\Bigl(\frac{y_{\text{t}} - t/n}{w_{y_{\text{t}}}/\sqrt{2}}\Bigr)^2
\biggr].\label{ETMt}
\end{eqnarray}
\end{widetext}
Replacing TM $\to$ TE and $\mu \to \epsilon$ in
(\ref{ETMir}) and (\ref{ETMt}) we obtain formulas for
$\mathcal{E}_{\text{TE}}(x,y,t)$. The free parameters are:
$\varphi_{\text{i}}$, $w_{x_{\text{i}}}$ and $w_{y_{\text{i}}}$.

In (\ref{ETMir}) and (\ref{ETMt}), one can recognize squares of the
envelopes for the fields (\ref{Hza}) and (\ref{Hzb}), which move along
the same axes and with the same velocities as the fields. The third term in
(\ref{ETMir}) describes the interference of the incident and reflected
pulses in vacuum. This term, different from zero when the incident and
reflected pulses overlap, is strongly oscillating in the laboratory
coordinate $y$, see Fig.~\ref{ydep}. The wavelength (of standing wave)
\begin{equation}
\lambda_{\text{sw}} =
\frac{1}{2\cos\varphi_{\text{i}}}\label{lamsw}
\end{equation}
depends on the angle of incidence $\varphi_{\text{i}}$ but is independent
of the pulse radii, $w_{x_{\text{i}}}$ and $w_{y_{\text{i}}}$.

Note that $w_{y_{\text{i}}}$ defines the laser pulse duration $\tau$.
Assuming that $\tau$ is defined as the width of the energy versus time
profile at half-maximum (for incident pulse), we obtain using (\ref{ETMir}),
\begin{equation}
\tau = (2 \ln 2)^{1/2} \, w_{y_{\text{i}}}
= 1.1774 \, w_{y_{\text{i}}}.
\label{tau}
\end{equation}

\section{Discussion and conclusions}\label{discus}

\begin{figure}
\includegraphics[scale=0.5]{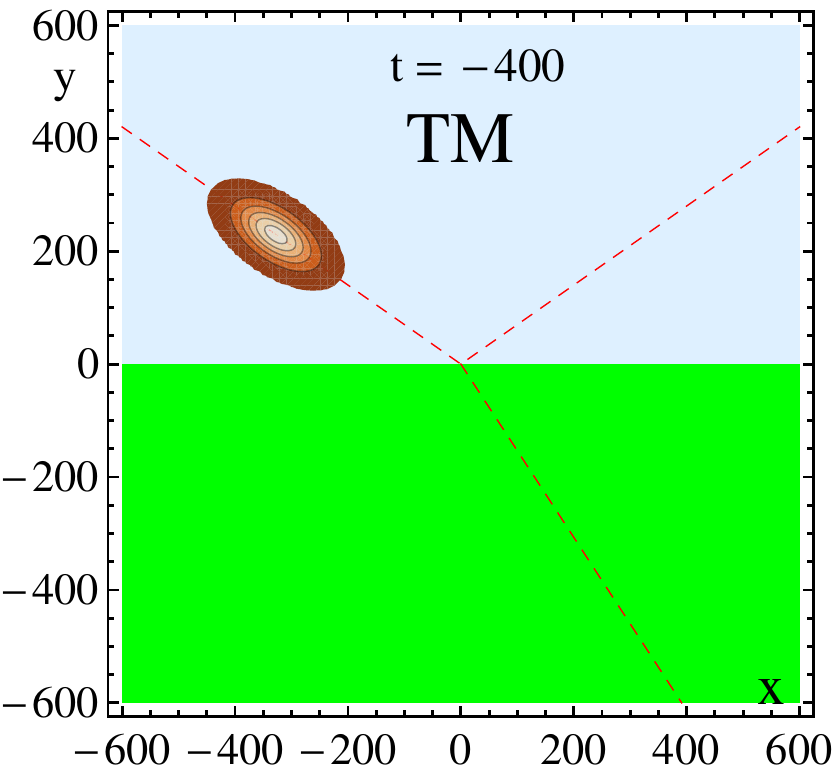}
\includegraphics[scale=0.5]{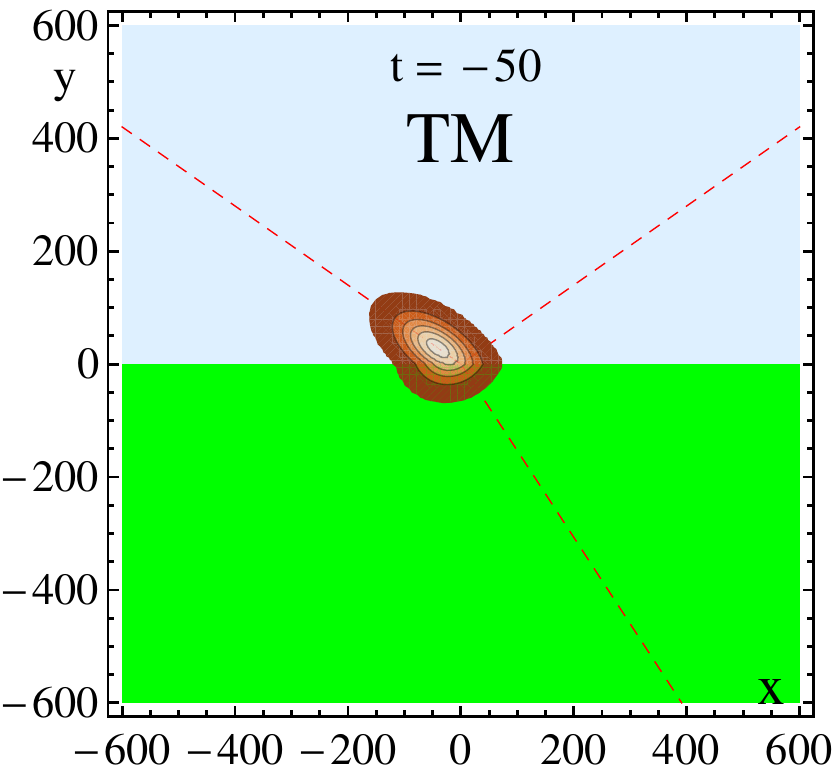}\\
\includegraphics[scale=0.5]{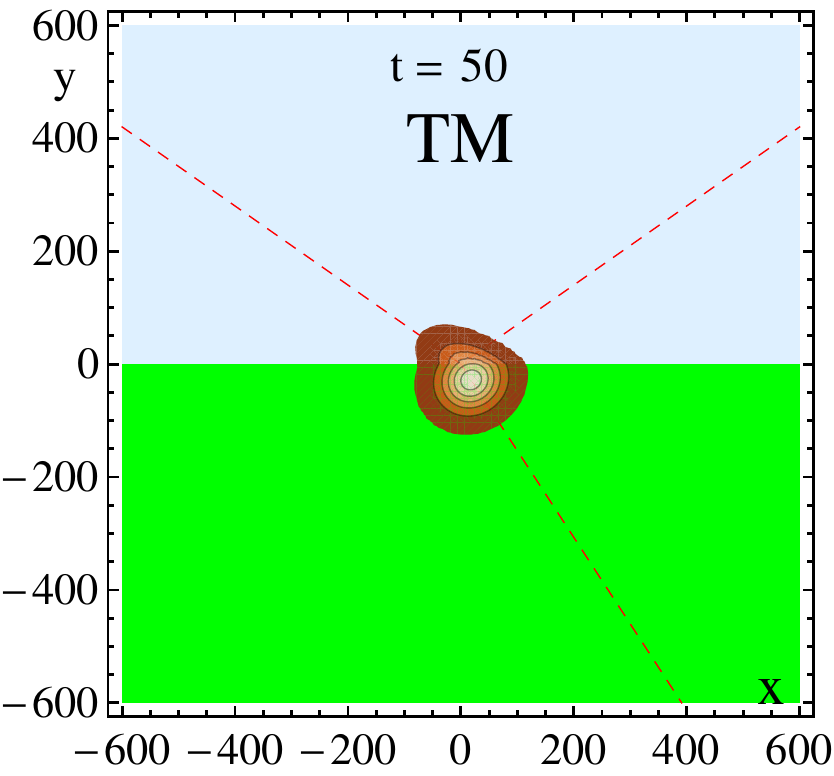}
\includegraphics[scale=0.5]{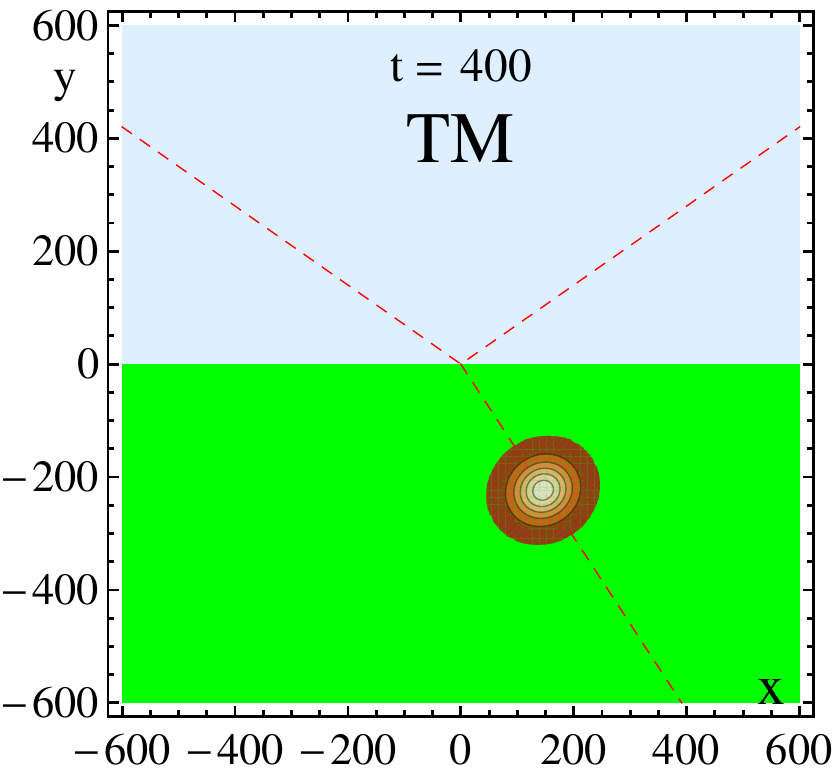}
\caption{(Color online) Results for $\mathcal{E}_{\text{TM}}
(x,y,t)$, spatially averaged over the interference peaks, for
$\epsilon = 2.25$, $\mu = 1$, $\varphi_{\text{i}} =
55^{\,\circ}$, $w_{x_{\text{i}}} = 50$, $w_{y_{\text{i}}} = 100$.
No noticeable reflection. The pertinent videos are given in \cite{vid1}
(our results as above) and in
\cite{nature} (experimental results of \cite{gao}).}
\label{resETM}
\end{figure}

\begin{figure}
\includegraphics[scale=0.5]{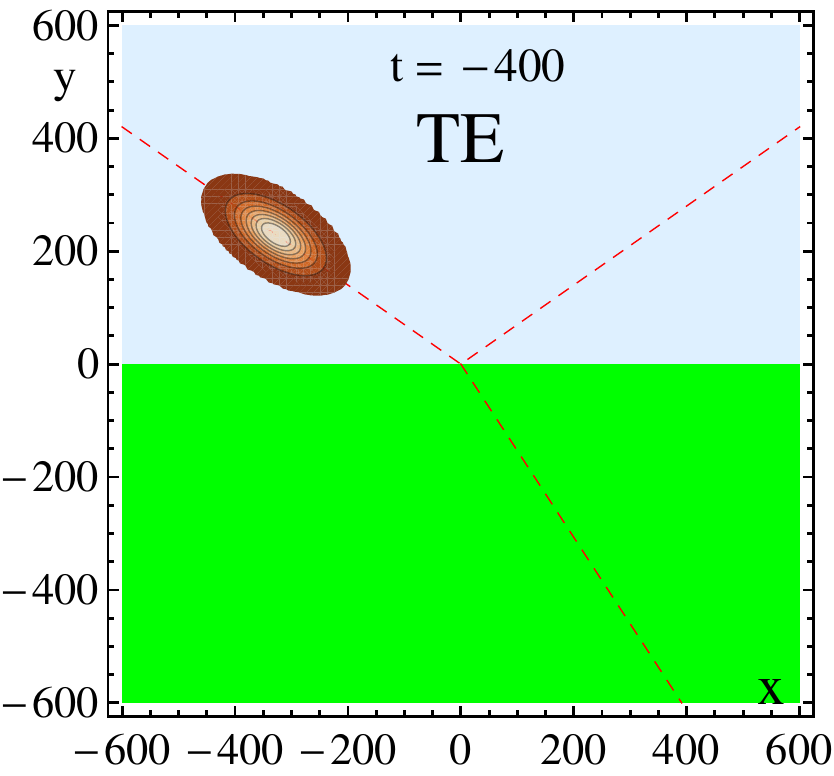}
\includegraphics[scale=0.5]{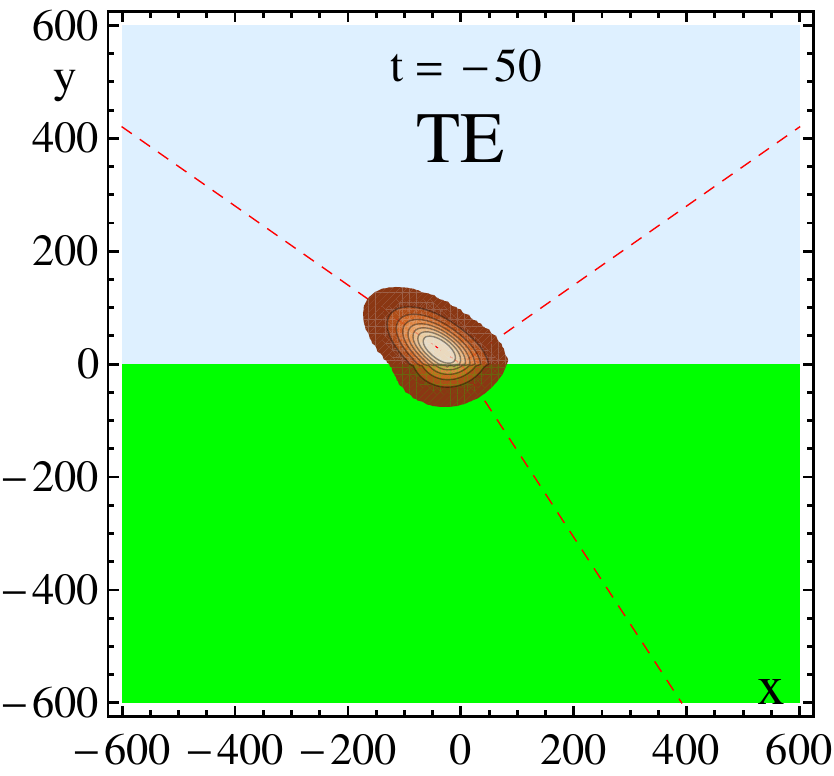}\\
\includegraphics[scale=0.5]{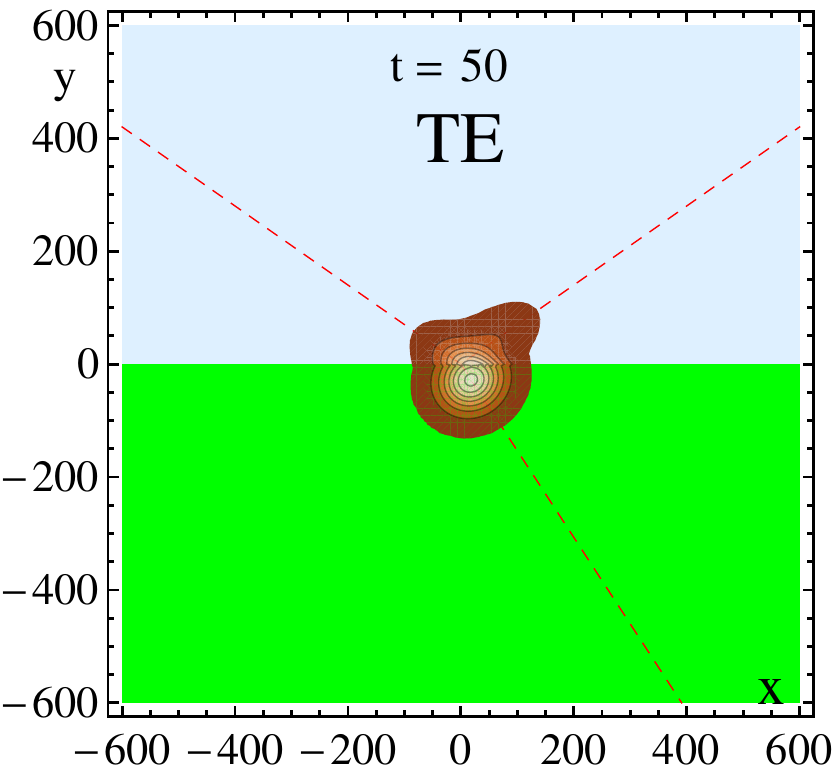}
\includegraphics[scale=0.5]{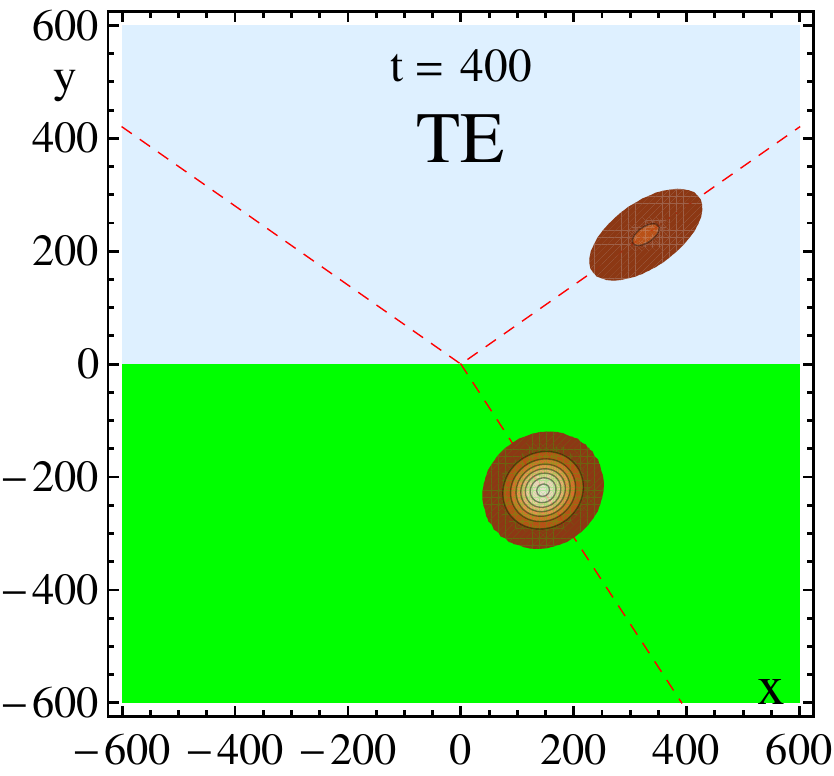}
\caption{(Color online) As in Fig.~\ref{resETM} but for
$\mathcal{E}_{\text{TE}}(x,y,t)$. See \cite{vid2} for the video.}
\label{resETE}
\end{figure}

The parameters used in \cite{gao} were: $n=1.5$ and $\mu = 1$, which
implied $\epsilon = 2.25$. Furthermore, $\varphi_{\text{i}} = 55^{\,\circ}$
and $w_{y_{\text{i}}}/w_{x_{\text{i}}} = 2$, as we could read from Fig.~3(b) in
\cite{gao}. The value of $\varphi_{\mathrm{i}}$ was close to the
Brewster angle $\varphi_{\mathrm{B}} \equiv \arctan n =
56.31^{\,\circ}$ for which $R_{\mathrm{TM}} = 0$
[$R_{\mathrm{TM}}^2(\varphi_{\mathrm{i}} = 55^{\,\circ})=1.77847
\times 10^{-4}$].
This could be the reason why no reflection was observed
in \cite{gao}, see Fig.~\ref{resETM}, and would suggest
the laser pulse to have TM polarization. In that case, by rotating the light
source by $90^{\,\circ}$ around the incident beam axis, the
TE polarization would be obtained, for which
$R_{\mathrm{TE}}^2(\varphi_{\mathrm{i}} = 55^{\,\circ}) = 0.1393$ versus
$\epsilon (1 + R_{\text{TE}})^2 = 0.8840$ for the refracted beam.
The reflected pulse should then be seen along with the incident and
refracted ones, see Fig.~\ref{resETE}.


In our calculations, based on the dimensionless formulas
(\ref{ETMir}) and (\ref{ETMt}), for convenience we have chosen the pulse
radii to be rather small:
$w_{x_{\text{i}}} = 50$ and $w_{y_{\text{i}}}=100$. The actual dimensions of
the pulses used in \cite{gao} were
much larger but, as we will see, the results in that case are closely
related to ours.
\begin{figure*}[t!]
\includegraphics[scale=0.44]{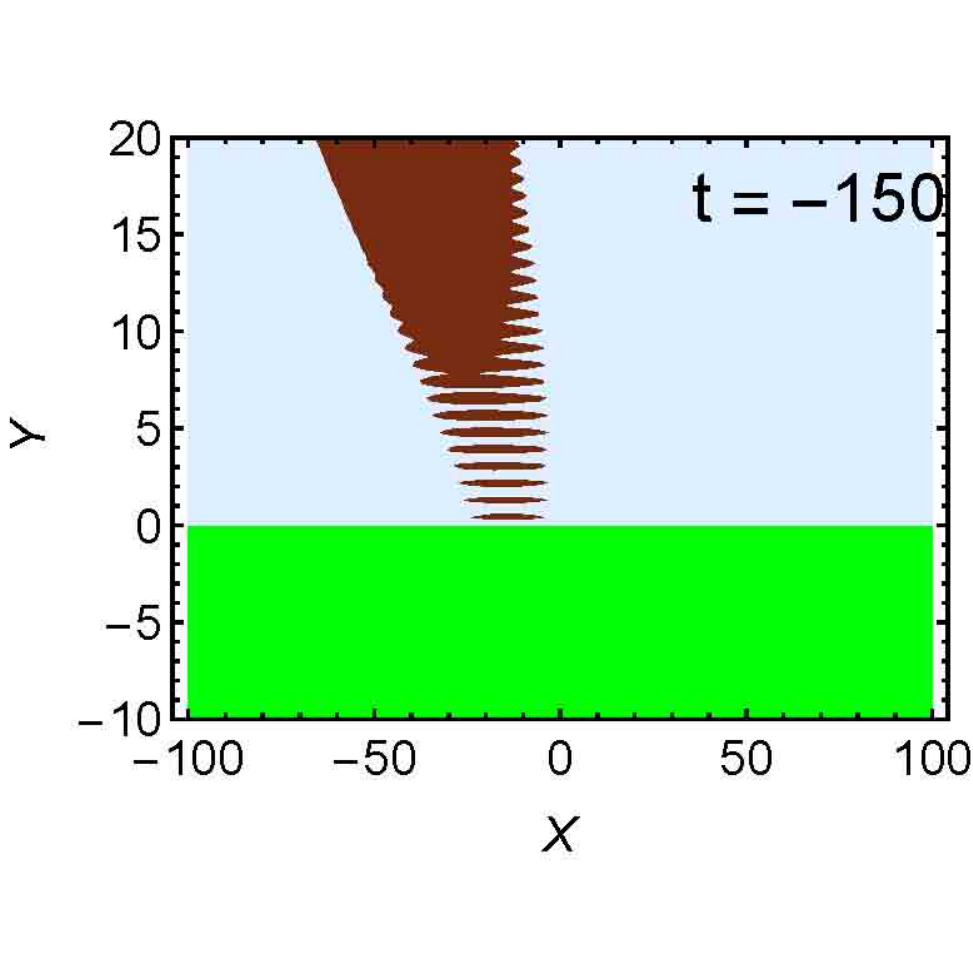}
\includegraphics[scale=0.44]{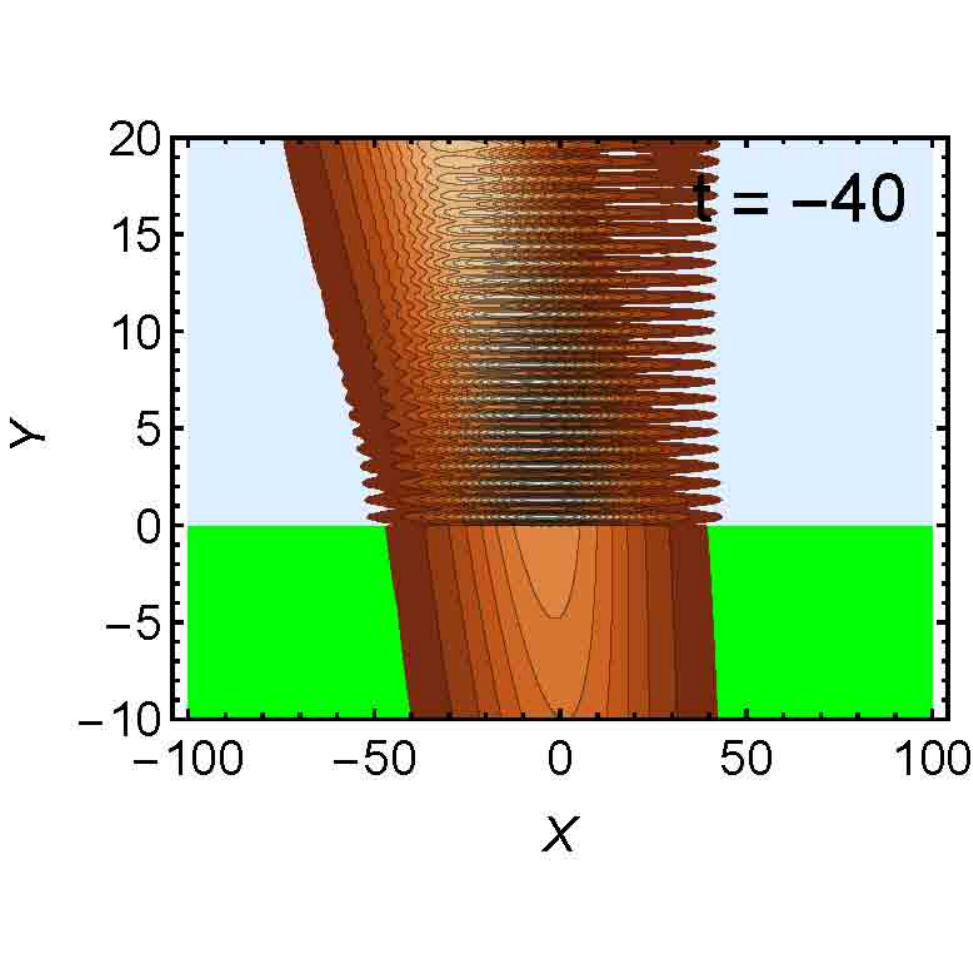}
\includegraphics[scale=0.44]{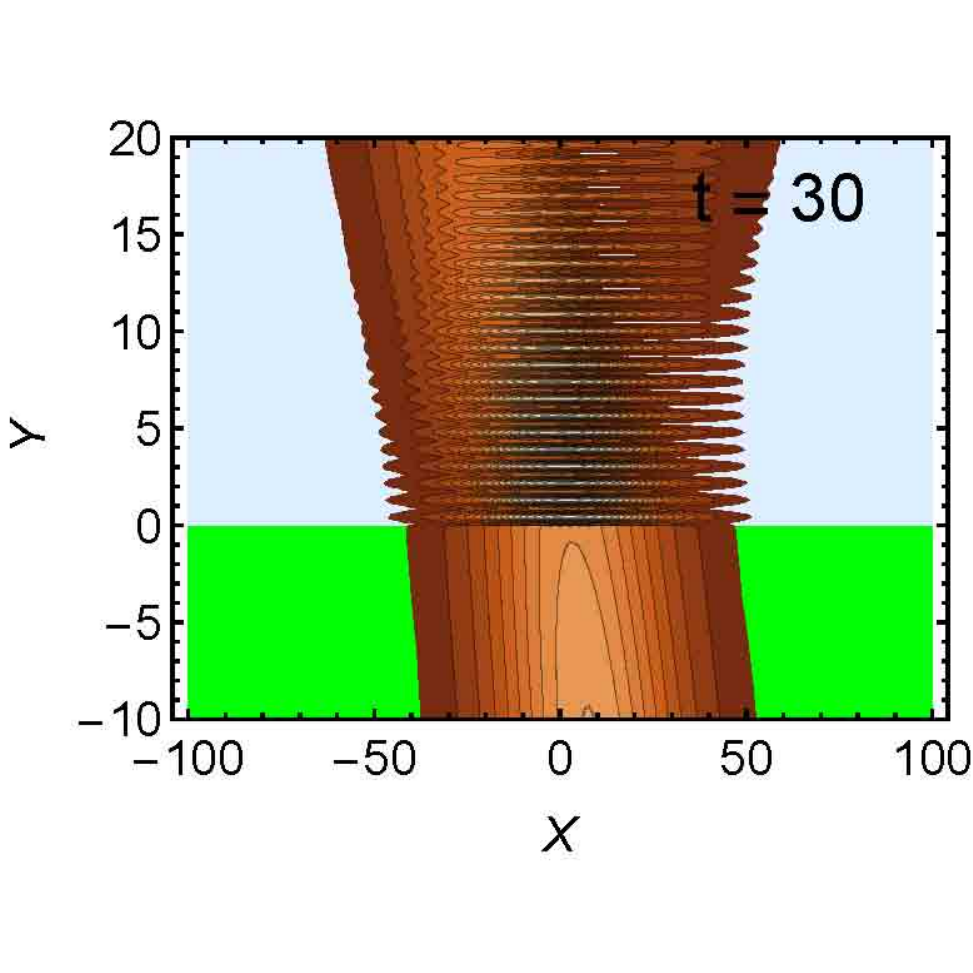}
\includegraphics[scale=0.44]{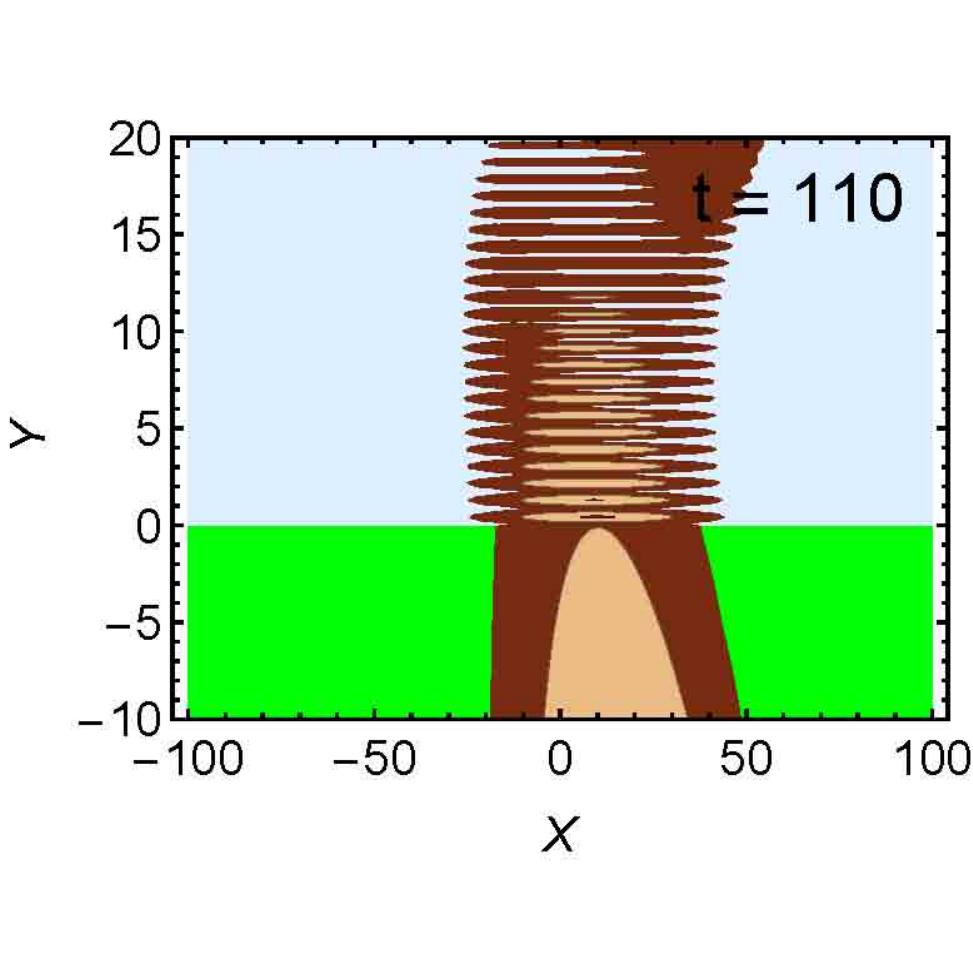}
\caption{(Color online) Results for $\mathcal{E}_{\text{TE}}
(x,y,t)$, $w_{x_{\text{i}}} = 20$, and
other parameters as in Fig~\ref{resETM}.}\label{intpeaks}
\end{figure*}

At any instant $t$, the energy densities $\mathcal{E}_{\text{TM}}(x,y,t)$
and $\mathcal{E}_{\text{TE}}(x,y,t)$
are functions of $x$ and $y$. They can be represented graphically by level
contours. If we multiply the pulse radii $w_{x_{\text{i}}}$ and
$w_{y_{\text{i}}}$ by some $a>0$, but at the same time multiply by $a$ also
$x$, $y$, and $t$, all exponents in (\ref{ETMir}) and (\ref{ETMt}) will
remain unchanged. Therefore, if furthermore we either drop the cos term
(i.e., replace it by its average value zero) or take the extreme values
($\pm 1$) of the cos, the contours in these three cases will scale along
with $a$. For example, they will be enlarged $a$ times if $a>1$. During this
scaling, the interference peaks shown in Fig.~\ref{ydep} will remain unchanged.
This scaling also means that the contours in question for $a \neq 1$ will be
the same as those for $a=1$, but one has to multiply by $a$ the numbers
associated with $t$ and with the $x$ and $y$ axes
(the interference peaks will then get squeezed $a$ times if $a>1$).
In particular, this scaling is applicable to the contours
shown in Figs.~\ref{resETM} and \ref{resETE}, where we present
``macroscopic'' results (averaged over the interference peaks by neglecting
the cos term).

Notice that for the incident beam, the LHS of (\ref{cond}) gains the factor
$a^{-2}$ if $y_{\text{i}}$ and $w_{x_{\text{i}}}$ are multiplied by $a$.
This increases the accuracy of (\ref{gaussb1}) if $a>1$. In our case ($a = 1$),
\begin{equation}
\lambda_{\text{sw}} = 0.8717, \qquad
\frac{y_{\text{i}}^2(t=-400)}{\hat{y}_{x{\text{i}}}^2} \approx
2.6 \times 10^{-3},\label{our}
\end{equation}
and the applicability condition (\ref{cond}) implies 
$a_{\text{min}} \simeq 0.1$ ($w_{x_{\text{i}}\,\text{min}} \simeq 5$).

The laser used in \cite{gao} was characterized by the wavelength
$\lambda_0=532$ nm  and the pulse duration $\tau = 7$ ps. Dividing this
$\tau$ by $T_0 = 177.33 \times 10^{-5}$ ps to make it dimensionless and
using (\ref{tau}) we obtain $w_{y_{\text{i}}}=3352.7$. This leads to
$a=w_{y_{\text{i}}}/100 \simeq 33.5$,
a number by which one has to multiply all numerical values in
Figs.~\ref{resETM} and \ref{resETE}, to make them applicable to the
experiment described in \cite{gao}.

In Fig.~\ref{intpeaks} we present the details of the interference peaks
evolution. The distance $\lambda_{\text{sw}}$ between the peaks is given
by (\ref{lamsw}) and (\ref{our}).

A strong validity test for our formulas and calculations
was a numerically confirmed continuity of the tangential components
of $\mathbf{E}$ and $\mathbf{H}$ across the boundary
($y=0^{\pm}$), and the fact that the total energy was conserved
(time independent):
\begin{equation}\label{acctest}
\int_{-1000}^{1000}dx\int_{-1000}^{1000}dy \,\mathcal{E}_{\text{TM}}(x,y,t)
 = 7853.982,
\end{equation}
for any $t \in [-400,400]$, and exactly the same result for
$\mathcal{E}_{\text{TE}}(x,y,t)$, see Figs.~\ref{resETM} and \ref{resETE}.
Incidentally, this conservation was also valid for
$\mathcal{E}_{\text{TM}}$ and $\mathcal{E}_{\text{TE}}$ spatially averaged
over the interference peaks, which can thus be treated as macroscopic
energy densities. Equation (\ref{acctest}) illustrates high accuracy of
the paraxial approximation in our application.

All calculations, figures, and videos were done by using \textit{Mathematica}.

\begin{acknowledgments}
The authors would like to thank E. Infeld, M. Rusek, S. Suckewer, and
P. Zi\'n for useful discussions.
\end{acknowledgments}


%

\end{document}